\documentclass[showpacs,preprintnumbers]{revtex4}
\usepackage{graphicx,color,psfrag}
\def\bc{\begin{center}}
\def\ec{\end{center}}

\def\beq{\begin{equation}}
\def\eeq{\end{equation}}

\def\bk{{\bf k}}
\def\bq{{\bf q}}

\begin{document}
\title{
Phase transitions in the two-dimensional electron-hole gas \\
}
\author{Oleg L. Berman$^{1,2}$, Roman Ya. Kezerashvili$^{1,3}$, and Klaus Ziegler$^{4}$}
\affiliation{\mbox{$^{1}$Physics Department, New York City College
of Technology, The City University of New York,} \\
Brooklyn, NY 11201, USA \\
\mbox{$^{2}$Kavli Institute for Theoretical Physics, UCSB, Santa
Barbara, CA 93106, USA} \\
\mbox{$^{3}$ The Graduate School and University Center, The
City University of New York,} \\
New York, NY 10016, USA \\
\mbox{$^{4}$   Institut f\"ur Physik, Universit\"at Augsburg\\
D-86135 Augsburg, Germany }}
\date{\today}

\begin{abstract}

A first order phase transition between a BCS phase and an insulating 
Mott phase for a gas of spatially separated electrons and holes with tunable Coulomb interaction
and variable density is predicted. In the framework of a BCS-like mean-field approach and a Landau expansion in terms
of the pairing order parameter the phase diagram is studied. This indicates several phases and phase transitions,
including an electron-hole plasma at low density and weak interaction, an intermediate BCS phase
with Cooper pairs and an electron-hole plasma at high density and weak interaction. The insulating
Mott phase appears for the strong interaction and low temperatures. We briefly discuss the possibilities
to realize these phases in realistic systems such as coupled quantum wells and graphene double
layers.
\vspace{0.1cm}


\end{abstract}

\maketitle








\section{Introduction}
\label{intro}

\begin{figure}[t]
\begin{center}
\includegraphics[width=8cm]{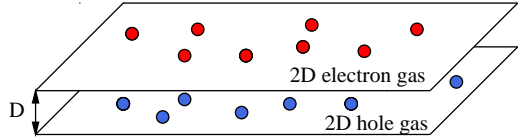}
\caption{
The formation of an electron-hole gas in two parallel layers separated by distance D, 
where the electrons reside in one, the holes in the other layer.
}
\label{fig:layers}
\end{center}
\end{figure}

An electron-hole gas, where the electrons reside in a two-dimensional layer
and the holes in another two-dimensional layer (see Fig. \ref{fig:layers}),
represents a many-body system with controllable attractive Coulomb
interaction and controllable density. For weak attraction the Bardeen-Cooper-Schrieffer 
(BCS) approach predicts the formation of Cooper pairs, whereas strong attraction
results in the formation of composite bosons, known as indirect excitons. 
Besides these two fundamental regimes other phases may exist. In particular, at
higher densities new correlated phases could appear when the Coulomb electron-hole
attraction is changed from weak to strong coupling through a change of the interlayer
separation. This could enable us to create an electron-hole plasma, a state of superfluid 
Cooper pairs and even a Mott state at high densities. The physics would go beyond
the weakly interacting electron gas in conventional superconductors, where
the attraction is created by electron-phonon coupling~\cite{Schrieffer}.

The double layer system can be realized in the form of semiconductor coupled
quantum wells (CQWs) or by two graphene layers (GLs). The CQWs are
conceptually simple: negative electrons are trapped in a two-dimensional
plane, while an equal number of positive holes is trapped in a parallel
plane a distance $D$.
Superfluidity in CQWs has been predicted by Lozovik and Yudson in 1975~\cite%
{Lozovik}, which stimulated intensive theoretical research for a number of
new physical phenomena such as persistent electric currents and unusual
effects in strong magnetic fields~\cite%
{Shevchenko,Lerner,Dzjubenko,Kallin,Knox,Yoshioka,Littlewood,Vignale,Ulloa}.
Moreover, it led to a series of experimental studies that have focused on
observing these effects~\cite%
{Snoke_paper,Snoke_paper_Sc,Chemla,Krivolapchuk,Timofeev,Zrenner,Sivan,Snoke,EM}%
. Other theoretical studies considered the BCS phase of electron-hole Cooper
pairs in a dense electron-hole system~\cite{Lozovik} and a dilute gas of
dipolar (indirect) excitons, where the latter are formed as bound states of
electron-hole pairs in CQWs~\cite{BLSC}.

Besides the condensation of excitons and the formation of a superfluid
state, their dissociation into the electron-hole plasma (EHP) is also of
interest and was studied experimentally for GaAs/AlGaAs CQWs in Ref.~%
[\onlinecite{Stern}]. The phase diagram of indirect excitons formed by the
spatially separated electrons and holes in GaAs/AlGaAs CQWs was analyzed
experimentally, and it was shown that the exciton system undergoes a phase
transition to an unbound electron-hole plasma for increasing temperatures ~%
\cite{Stern}. This transition has been manifested as an abrupt change in the
photoluminescence linewidth and in the peak energy at some critical power
density and temperature. The dynamics of the ionization transition of the
excitons was studied by using the rate equation~\cite{Snoke_pre}. The
fraction of ionized carriers in an electron-hole-exciton gas in a
photoexcited semiconductor was derived theoretically by applying the
mass-action equation, or Saha equation, for the number of free carriers in
equilibrium~\cite{Snoke_SSC}.

Within the Hartree-Fock approximation the ionization equilibrium of an
electron-hole plasma and the exciton phase was investigated in a highly
excited semiconductor in Ref.~[\onlinecite{Manzke_PRB}] with special
attention to the influence of many-particle effects such as screening and
lowering of the ionization energy. The dissociation of excitons in
GaAs-GaAlAs quantum wells with increasing excitation was studied within two
different theoretical approaches~\cite{Manzke_NJP}. From a thermodynamic
approach a simple criterion for the transition to the EHP was obtained: the
sum of chemical potentials of carriers, reflecting the effective shrinkage
of the band edge, crosses the exciton energy with increasing excitation.
Alternatively, a spectral approach, based on the semiconductor Bloch
equations within linear optical response, was used to analyze the
quasi-particle properties of carriers and the dynamical screening between
electron-hole pairs. While the first is effectively a one-particle approach,
in the second the whole two-particle spectrum is studied~\cite{Manzke_NJP}.
The transition between the excitons 
and the EHP was obtained in Hartree-Fock approximation for QWs~\cite%
{Manzke_NJP}. This transition was also studied using the thermodynamic
perturbation theory~\cite{laikhtman03}. 

The condensation of Cooper pairs formed by spatially separated electrons and
holes was also the subject of more recent studies in a system of two
isolated graphene layers \cite{Pesin,Ogarkov}. The electron-electron
interactions in decoupled GLs were analyzed in detail, where it was found
that the Hartree-Fock approach provides a quite accurate description of the
inter-particle interactions in GLs~\cite{Polini}. The electron-hole
superfluidity caused by formation of the BCS phase by Cooper electron-hole
pairs in two parallel bilayer graphene sheets was proposed recently~\cite%
{Perali}, where also the Hartree-Fock approximation was applied. In Ref. 
\cite{berman12} was predicted the superfluidity of quasi-two-dimensional
dipole excitons in double-layer graphene in the presence of band gaps.

According to a detailed study of the ionization degree of the electron-hole
plasma in semiconductor quantum wells~\cite{Portnoi}, the dependence of the
degree of ionization on the carrier density depends dramatically on the
electron and hole masses and the dielectric constant, which are different
for different semiconductors. In Ref.~[\onlinecite{Portnoi}] the phase
transition between the exciton phase and the EHP was analyzed. Finally, the
transition from an exciton gas to an EHP could have important applications
to all-optical switching~\cite{Steger,Snoke_nn,Ballarini}.

In this Paper we predict a first oder phase transition between the BCS phase and an
insulating Mott phase for strong electron-hole attraction at low
temperatures for the electron-hole gas in the CQWs when one filled with
electrons, the other with holes. This transition can be observed  at the
small interlayer separation $D$ and high electron and hole densities. We
study the phase diagram of the electron-hole gas formed in two layers within
a mean-field approximation for the pairing order parameter as a function of
interaction strength, temperature and density. The calculation includes a
Landau expansion in terms of the pairing order parameter. 
The analysis of the phase diagram indicates several phases and phase transitions, 
including an electron-hole plasma at low density and weak interaction, an intermediate BCS phase with Cooper pairs 
and an electron-hole plasma at high density and weak interaction.
For realistic systems of CQWs and two GLs we consider the case where the masses of the
electrons and holes are different. This difference 
results in the extra term in the BCS equation for the free energy. Besides,
we analyze the order of the phase transitions with respect to the
temperature as well as the temperature dependence of the order parameter,
using the results of the mean-field calculations at nonzero temperatures. 


The paper is organized in the following way. In Sec.~\ref{modham} the model
Hamiltonian of two-layer electron-hole system is introduced. The mean-field
approximation, applied to the system under consideration, is described in
Sec.~\ref{mfa}. The phase diagram of a translational invariant electron-hole
system is presented in Sec.~\ref{pdt}. Finally, the discussion of the
results and conclusions follow in Sections~\ref{disc} and \ref{conc},
respectively.

\section{Model Hamiltonian}

\label{modham}

The Hamiltonian of a spinless electron-hole gas in momentum representation can be written as
\beq
H=\sum_{\bk}\sum_{\sigma=e,h}(\epsilon_{\bk,\sigma}-\mu_\sigma)c^\dagger_{\bk\sigma}c_{\bk\sigma}
-\sum_{\bk,\bq,\bq'}
U_\bk
c^\dagger_{\bk-\bq,h}c_{\bk-\bq',h}c^\dagger_{\bq,e}c_{\bq',e}
\ , \label{hamilton} \eeq where $c^\dagger_{k,e}$ ($c_{k,e}$) is the
creation (annihilation)  operator for electrons, and
$c^\dagger_{k,h}$ ($c_{k,h}$) is the corresponding operator for
holes,  $\mu_\sigma$ is the chemical potential of fermions, assuming
that they are adjusted in such a way that the  densities of
electrons and holes are equal in order to have a neutral
electron-hole plasma, which is justified by the fact that the
electrons and holes are created always pairwise by an external Laser
source. The spin of electrons and holes is neglected here because we
are not interested in magnetization effects. Moreover, we assume the
simple parabolic dispersion relations
$\epsilon_{\mathbf{k},e}=k^2/2m_{e}$,
$\epsilon_{\mathbf{k},h}=k^2/2m_{h}$ for the electrons and holes
with electron $m_e$ and hole $m_h$ masses, respectively. To take
care of the band structure of a realistic system we introduce an
upper cut-off energy $E_b$ in such way that the kinetic energy of the
quasiparticles is constrained by
$0\le\epsilon_{\mathbf{k},e}+\epsilon_{\mathbf{k},h}\le E_b$.
 In Eq. (\ref{hamilton})
$U_{\bk}$ is the attractive electron-hole interaction, that is
different for the paired electrons and holes in CQWa and  two
separated graphene layers and can be defined by the 2D Fourier image
of the screened electron-hole attraction $U(\mathbf{r})$ as
\cite{Lozovik}
\begin{equation}
U_{\bk}= -\frac{\hbar \bar{U}\exp \left( -kD/\hbar\right)
}{k+2\hbar\left( a_{e}^{-1}+a_{h}^{-1}\right) +4\hbar\left( 1-\exp
\left( -2kD/\hbar\right) \right) /(a_{e}a_{h}k)}  \ . \label{2}
\end{equation}
In Eq.~(\ref{2}) ${\bar{U}}=2\pi \kappa e^{2}/(\varepsilon
l_{F}^{2})$ is the interaction strength, $\varepsilon $ is the
dielectric constant of the CQWs or the dielectric between GLs,
$\kappa =9\times 10^{9}\mathrm{N m^{2}/C^{2}}$, $e$ is the electron
charge, $a_{e,h}=\hbar ^{2}\varepsilon /(m_{e,h}e^{2})$, and $D$ is
the thickness of the dielectric interlayer.

Using the notation $E_{F}$ and $q_{F}$ for the Fermi energy
and the Fermi momentum, respectively, the Fermi momentum of the 2D Fermi
system is defined as $q_{F} = \hbar \sqrt{\pi n}$, where $n$
is the 2D density of electrons and holes. Estimating $q_{F}D$ for
the parameters of the real quantum wells, we obtain $q_{F}D/\hbar
\ll 1$, and we use the following approximation for the potential
energy of the electron-hole attraction given by Eq.~(\ref{2}):
\begin{equation}
U_{\bk}\approx -\frac{\hbar \bar{U}e^{-kD/\hbar }}{k+a} \ ,
\label{static_screening}
\end{equation}
where $a$ is a material dependent parameter (cf. Table \ref{table:1}) given by
\begin{eqnarray}
\label{LY} a = 2 \varepsilon \hbar \left(a_{e}^{-1} + a_{h}^{-1} +
4Da_{e}^{-1}a_{h}^{-1}\right) \ .
\end{eqnarray}


Now we consider the grand-canonical equilibrium of the electron-hole
gas, and define the expectation value of the operator $O$ with
respect to the Boltzmann distribution $\exp(-\beta H)$ at
temperature $T$ and the trace with respect to all fermionic quantum
states of the Hamiltonian $H$  as: \beq \langle O \rangle={1\over Z}
Tr\Big[ e^{-\beta H} O\Big] , \ \ \ Z=Tr\Big[ e^{-\beta H}\Big] \ ,
\label{partition00} \eeq where $\beta=1/k_B T$. The free energy $F$
then reads $F=-k_BT\log Z$, from which we obtain the densities of
electrons and holes as  \beq \langle n_{e,h}\rangle =-\frac{\partial
F}{\partial\mu_{e,h}} \ , \label{density00} \eeq where $\mu_{e}$ and
$\mu_{h}$ are the chemical potentials of the electron and hole
gases, respectively.

\section{Mean-field approximation}

\label{mfa}

In order to evaluate the free energy of
the interacting electron-hole gas we apply a mean-field
approximation. This is based on the idea that we replace the
interaction terms in the Hamiltonian (\ref{hamilton}), which are
quartic expression with respect to the fermion operators, by a
quadratic term that couples to a mean field. There are several
options for choosing the mean field. Very common
is to use a Hartree-Fock like approximation that couples to the
fermion densities $c^\dagger_{\bk,\sigma}c_{\bk,\sigma}$
\cite{Manzke_PRB}. Then the mean field is a self-energy
$i\Sigma''_{\bk,\sigma}+\Sigma'_{\bk,\sigma}$, where the real part
$\Sigma'_{\bk,\sigma}$ describes a shift of the Fermi energy and the
imaginary part $\Sigma''_{\bk,\sigma}$ a damping of the quantum
dynamics. An alternative choice
is to consider a complex mean field $\Delta_\bk$ that corresponds to
BCS pairing \cite{Lozovik}: \beq \sum_{\bk,\bq,\bq'} U_\bk
c^\dagger_{\bk-\bq,h}c_{\bk-\bq',h}c^\dagger_{\bq,e}c_{\bq',e}
\approx \sum_{\bk}\Delta_\bk c_{\bk,e}c_{-\bk,h} +h.c. \ ,
\label{mfa1} \eeq where the mean field $\Delta_\bk$ is the usual gap
order parameter of the BCS theory. In the latter case we can assume that
the effective shift of the Fermi energy by the self-energy is already
taken into account, which implies that the Fermi energy is the renormalized
one.
Within this approximation we can perform the trace in Eq.
(\ref{partition00}) with respect to non-interacting electrons and
holes. These states are created as product states by applying the
fermionic creation and annihilation operators $c^\dagger_{\bk,e}$
($c_{\bk,e}$) and $c^\dagger_{\bk,h}$ ($c_{\bk,h}$) of Eq.
(\ref{hamilton}) to a vacuum state. Then the free energy in momentum
representation reads (cf. Ref. \cite{ziegler05}) 
\beq F
=-\frac{1}{(2\pi q_F)^2}\left[ \int_{q\le q_F} \frac{\Delta_\bq
\Delta_{-\bq}}{U_\bq}d^2q
+
\int_{q^2/2m_+\le E_b}\log\Big(1+e^{-2\beta E_q 
}
+2e^{-\beta E_q 
}\cosh\left[\beta\sqrt{RE_q^2 +|\Delta_\bq|^2}\right]\Big) d^2q
\right] \ , \label{mft1} 
\eeq 
where $E_q=(q^2-q_F^2)/2m_+$ is the
single particle energy with $m_\pm=m_e m_h/(m_h\pm m_e)$ and $
R=m_+^2/m_-^2=(m_h-m_e)^2/(m_h+m_e)^2 $. 
It should be noticed that we consider the case where the masses of the 
electrons and holes are different. This results in the extra term 
related to R in Eq. (8). When the effective masses of the electrons 
and holes are equal this term vanish. Now the gap order parameter
$\Delta_\bq$ is determined as the minimum of the free energy. The
density of particles can be evaluated using Eq.~(\ref{density00}),
which allows us to fix
the Fermi energy $E_{F(e,h)}=q_F^2/2m_{e,h}$ 
by adjusting $\langle n\rangle$ with the experimental results. This
allow us to calculate first the pairing order parameter $\Delta_\bq$
as the minimum of the free energy, after inserting the solution in
$F$, then the average density $\langle n\rangle$.

By increasing the laser pumping intensity, the electron-hole density
increases and one can observe a phase transition from a dilute
electron-hole plasma to a BCS (paired electron-hole system) state
and another phase transition from the BCS state to dense EHP.  Both
phase transitions are the second order transition. Besides the
transition from the BCS phase to the EHP we would also expect a
transition similar to the Mott transition in a repulsive Hubbard
model. This is due to the fact that there is always competition
between kinetic energy and interaction energy. In the strongly
interating regime the interaction suppresses the kinetic energy and
an insulating Mott phase appears \cite{mott68}. A similar effect is
expected for our electron-hole gas in the strongly interacting
regime.

Before we provide a more detailed mean-field calculation, a
brief qualitative discussion of the electron-hole gas is given. The
properties of the system are controlled by three parameters, namely
the Fermi energy, the interaction strength and the
temperature. The influence of these parameters one can
understand in terms of the free energy in Eq. (\ref{mft1}) as follows:
(I) Changing the Fermi energy $E_F$ results
in a change of the density. (II) An increase of the temperature
implies an increase of the thermal fluctuations of the electron-hole
gas. (III) Finally, increasing the interaction (e.g., by reducing
the distance $D$ between the layers), will reduce the size of the
Cooper pairs in the BCS regime: With the screened Coulomb
interaction in (\ref{static_screening}) an increasing interaction
means a decreasing size of Cooper pairs, namely the appearance of
Cooper pairs with larger values of $q$. This is easy to justify by
using the linearized mean-field equation $\delta F/\delta\Delta_\bq=0$
for small $\Delta_\bq$ that gives
\beq
\Delta_\bq\sim 
f_1\frac{\hbar \bar{U}e^{-qD/\hbar }}{q+a}\Delta_\bq
\eeq
with a positive prefactor $f_1$, which is independent of ${\bar U}$.
This equation has a nonzero solution (i.e., a Cooper pair) if the factor
in front of $\Delta_\bq $ on the right-hand side is 1.
For large values of ${\bar U}$ this is the case only for a sufficiently
large value of $q$. In other words, for the strongly interacting regime
the Cooper pairs become more like tightly bound electron-hole pairs,
which can be considered as hard-core bosons. The hard core is
a consequence of the Pauli principle. This picture suggests to
present the electron-hole gas for weak interaction in terms of a
mean-field theory and for strong interaction as a hard-core Bose
gas. Below we focus on the mean-field theory alone, using Eq.~(\ref{mft1})
for the free energy and study the crossover to strong interaction
and high densities within this approach.

\section{Phase diagram of a translational invariant system}

\label{pdt}

The free energy expression (\ref{mft1}) is now considered for a
translational invariant system with weak interaction. In this case
we expect that the order parameter is uniform, i.e. is independent
on the momentum: $\Delta_{\mathbf{q}} = \Delta$.  This means that
the Cooper pairs are extended object, which may be justified at
least for weak interaction $u_0\ll E_F$, where $u_{0}$ is defined as
the Fourier component of the inverse Coulomb interaction
$1/u_0=\int_{q\le q_F} (1/U_\bq)d^2q/(2\pi q_F)^2$. Then the free
energy in Eq. (\ref{mft1}) reduces, up to a $\gamma$ independent
term, to \beq F=\frac{(k_BT)^2}{u_0}\left[
\gamma^2-\frac{u_0}{E_F}\frac{1}{4\pi}\int_{-E_F/k_BT}^{(E_b-E_F)/k_BT}\log\left(
1+\frac{\cosh\left(\sqrt{Ry^2+\gamma^2}\right)-1}{2\cosh^2(y/2)}
\right)dy\right] \label{mft2} \eeq with the dimensionless order
parameter $\gamma=\beta|\Delta|$. The integrand in Eq. (\ref{mft2})
is a symmetric function of $y\equiv E_q-E_F/k_B T$ which implies
that the integral is a symmetric function of $E_F$ with respect to
$E_F=E_b/2$. This property reflects the particle-hole symmetry of
the underlying fermionic Hamiltonian.

\begin{figure}[t]
\begin{center}
\includegraphics[width=8.5cm,height=7cm]{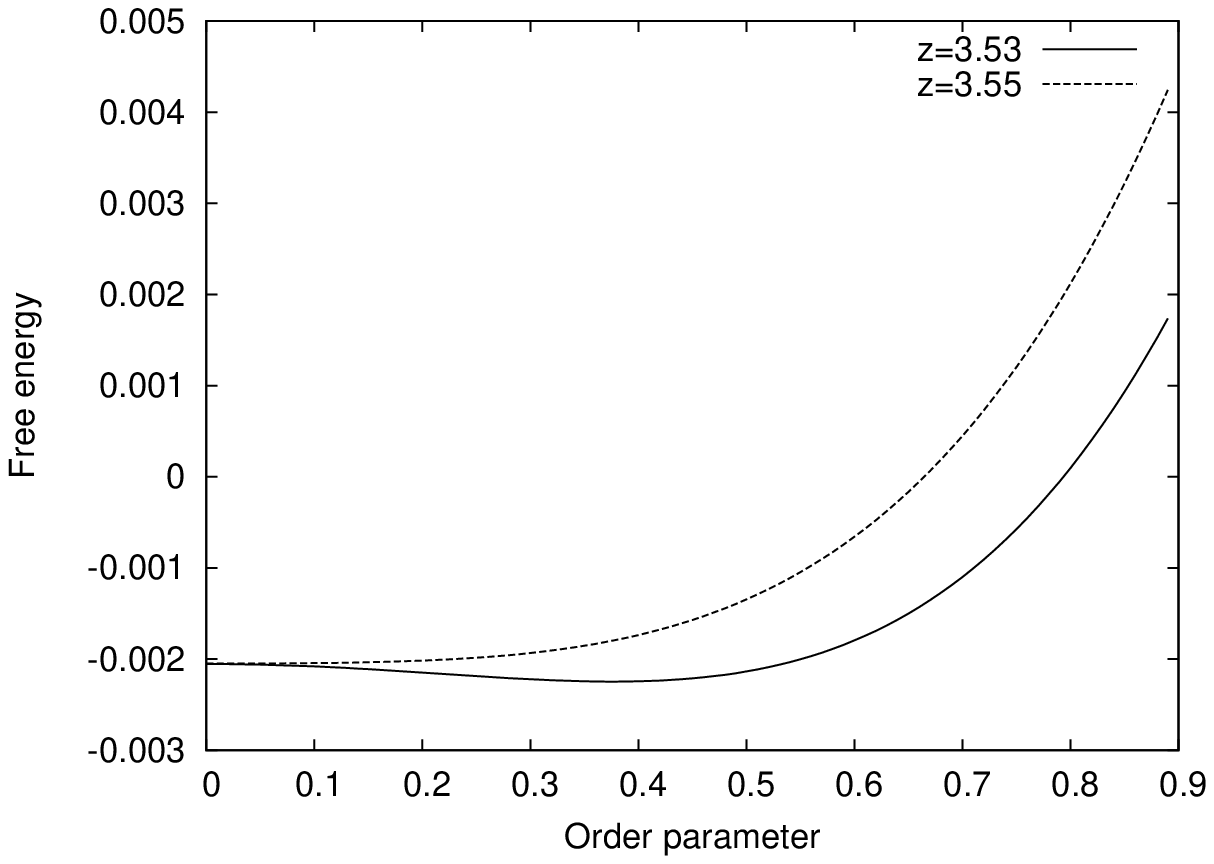}
\includegraphics[width=8.5cm,height=7cm]{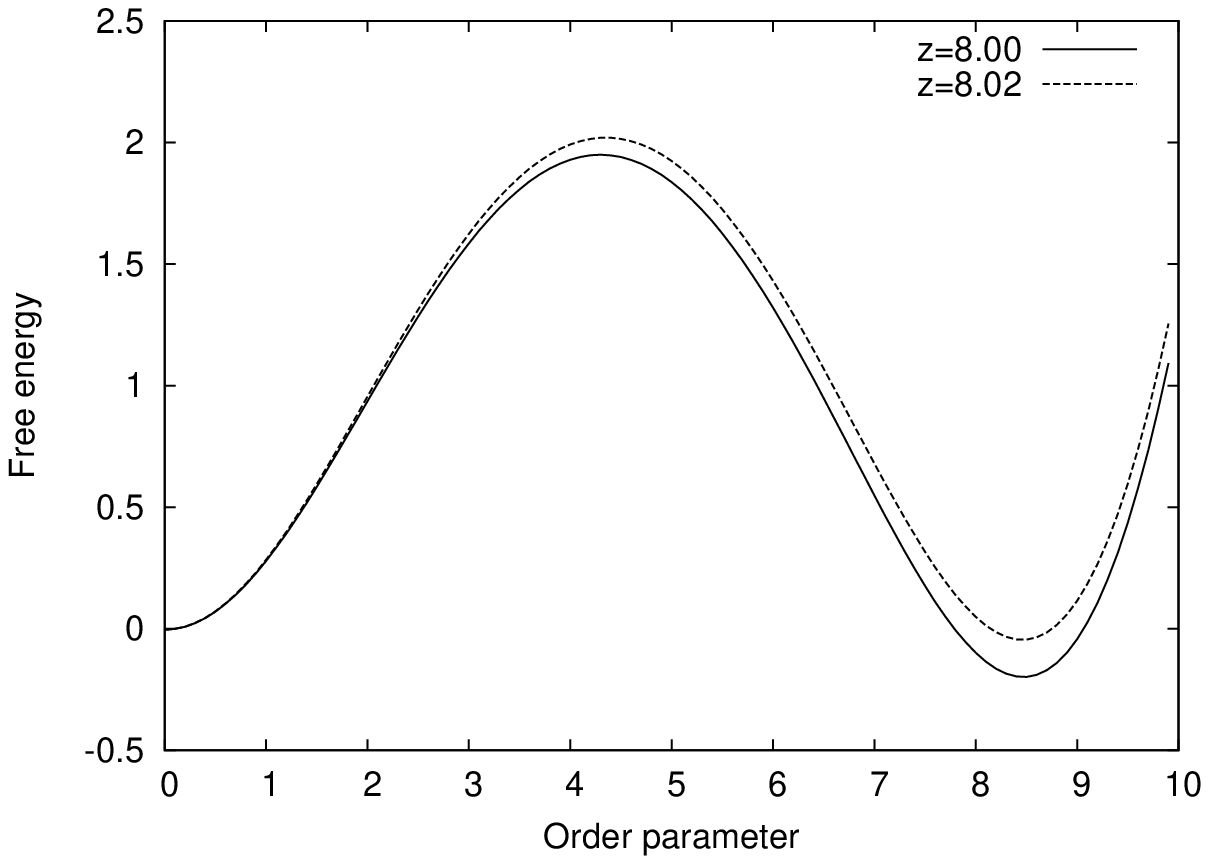}
\caption{
The free energy of Eq. (\ref{mft2}), normalized by $(k_B T)^2/u_0$, as a function of 
the dimensionless order parameter $\gamma$ and $z=E_F/k_B T$ for the upper band edge $
E_b=7 k_B T$. It indicates a second order phase transition for weak interaction 
($u_0=4.8 \pi E_F$) at $z=3.54$ (upper panel) and a first order phase transition
for strong interaction ($u_0=10.4 \pi E_F$) at $z=8.01$ (lower panel).
}
\label{fig:energy}
\end{center}
\end{figure}

Below  the results of the calculations for the electron-hole pairs
in GaAs/AlGaAs quantum wells are presented. As an example we have
plotted in Fig. \ref{fig:energy} the free energy as a function of
$\gamma$ and $E_F/k_BT$ for specific values $E_F/u_0$ and $E_b$. The
analysis of the results presented in Fig.~\ref{fig:energy} shows
that there is apparently a second order transition at $u_0=4.8\pi
E_F$ and a first order transition for stronger Coulomb interaction
$u_0=10.4\pi E_F$ from the BCS state at $E_F<E_c$ to a state with
vanishing order parameter.

\begin{figure}[t]
\begin{center}
\psfrag{E_F/k_B T}{$E_F/k_B T$}
\psfrag{F_4, F_6}{Coefficients $F_4$, $F_6$}
\psfrag{F_4}{$F_4$}
\psfrag{F_6}{$F_6$}
\includegraphics[width=8.5cm,height=7cm]{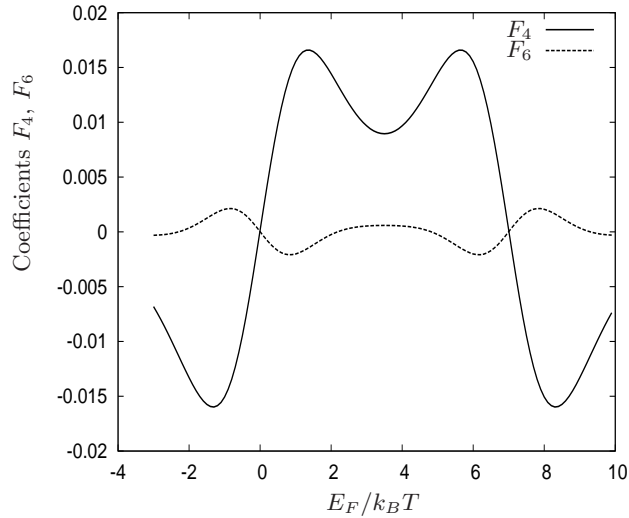}
\caption{
The Landau coefficients $F_4$ and $F_6$ of the free energy of Eq. (\ref{truncated}) as a function of $E_F/k_B T$
for $E_b=7 k_B T$.
They are symmetric with respect to the band center $E_F=E_b/2$.
}
\label{fig:coefficients}
\end{center}
\end{figure}

At lower densities there is a conventional BCS transition from a
low-density EHP to a superfluid phase. In order to describe the
phase transition scenario in more detail, we consider the case where
the order parameter is small, which is a realistic situation. Thus,
we can expand the free energy and present it in the Landau form as
\beq F(\Delta)=F_0+F_2\gamma^2+F_4\gamma^4+F_6\gamma^6 + O(\gamma^8)
\ , \label{truncated} \eeq where we call the polynomial up to
$\gamma^6$ the truncated free energy ${\cal F}_6(\gamma)$. The
latter allows us to discuss the phase diagram of the electron-hole
gas in a simple manner. First of all, the stability of the system
requires $F_6>0$. Then there are two different phases due to minima
of truncated free energy, namely one for $\gamma>0$ (superfluid
phase) and another one for $\gamma=0$  (electron-hole plasma). There
are two different types of a transition between these phases: a
first order and a second order transition.  They are distinguished
by the fact that the position of the minima $\gamma_0$ goes
continuously (second order transition) or discontinuously (first
order transition) to zero. In terms of our truncated free energy
${\cal F}_6(\gamma)$, the second order transition is characterized by one
minimum at $\gamma=0$, which is realized for $F_2>0$, and by a
maximum at $\gamma=0$ and another minimum at $\gamma>0$. The latter
is realized for $F_4>0$ and $F_2<0$. The first order transition, on
the other hand, is characterized by two minima and a maximum in
between. This is realized for $F_4<0$, $F_2, F_6>0$. The transition
point is given by the degeneracy of the two minima of ${\cal
F}_6(\gamma)$, namely for ${\cal F}_6(\gamma_+)={\cal F}_6(0)=F_0$
with \beq
\gamma_+^2=-\frac{F_4}{3F_6}+\sqrt{\frac{F_4^2}{9F_6^2}-\frac{F_2}{3F_6}}
\ . \label{op2} \eeq The coefficients $F_4$ and $F_6$ are plotted as
functions of the Fermi energy in Fig. \ref{fig:coefficients},
showing the particle-hole symmetry with respect to the band center
$E_F=E_b/2$.

A closer inspection of condition ${\cal F}_6(\gamma_+)={\cal F}_6(0)$ leads to 
the condition (cf. App. \ref{app:1})
\beq
F_4^2=4F_2 F_6
\eeq
for the phase transition. This equation and the condition for a second order 
phase transition $F_2=0$, $F_4>0$ shall be used in the subsequent section 
to evaluate the phase diagram of the electron-hole gas.

The phase diagram of the electron-hole gas within mean-field
approximation is depicted in Fig. \ref{fig:phasediagram}, which
clearly shows a paired BCS phase under the dome and a non-BCS phase
above the dome. This structure indicates that for a sufficiently
strong interaction at intermediate densities the electron-hole gas
has a tendency to form paired states. At lower or at higher
densities there is no pairing effect. The horizontal line at
$E_F/u_0\approx 0.04$ is important. It separates the strongly
interacting regime from the weakly interacting regime, which can be
distinguished by the fact that the former has a first order
transition to the non-BCS state and the latter has a second order
transition (see also Fig. \ref{fig:energy}). If the pairing order
parameter vanishes, the electron-hole gas is just a gas of
non-interacting electrons and holes with the Hamiltonian in the
mean-field approximation:
$H_0=\sum_{\bk}\sum_{\sigma=e,h}(\epsilon_{\bk,\sigma}-\mu_\sigma)c^\dagger_{\bk\sigma}c_{\bk\sigma}$.
The interaction is taken care of by a renormalized chemical
potential $\mu_\sigma$. If the Fermi energy $E_F=\mu_e+\mu_h$ is
inside the band (i.e., for $0<E_F$, $E_F<E_b$) this describes an
electron-hole plasma. On the other hand, if the Fermi energy is
above the band (i.e, for $E_F>E_b$) there are no extended (Bloch)
states and the system is in a localized or insulating phase. Using
here the analogy with the hard-core Bose gas, this could be a Mott
state of hard-core bosons \cite{moseley08}. However, this regime
cannot be described correctly by our mean-field theory.

\section{Discussion}

\label{disc}

\begin{figure}[t]
\begin{center}
\psfrag{E_F/k_B T}{$E_F/k_B T$}
\psfrag{E_F/U}{$E_F/u_o$}
\psfrag{first order transition}{$1^{\rm st}$ order transition}
\psfrag{second order transition}{$2^{\rm nd}$ order transition}
\includegraphics[width=8.5cm,height=7cm]{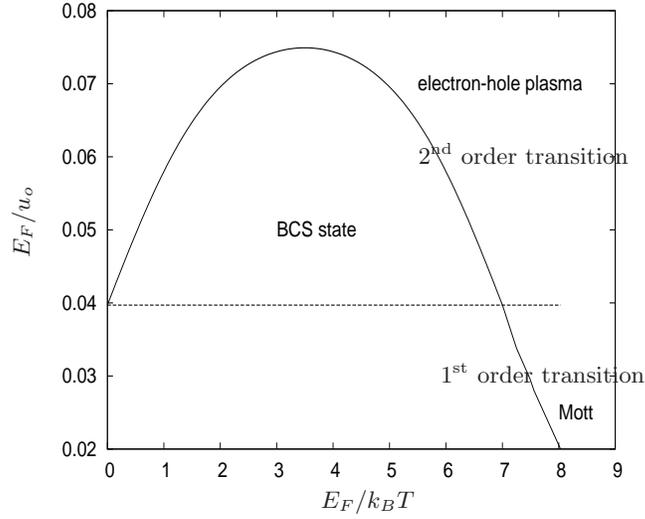}
\caption{
Phase diagram of the electron-hole gas with attractive interaction.
The phase boundary continuous symmetrically to negative values of the Fermi energy (not plotted here).
This is a direct consequence of the symmetric coefficients in Fig. \ref{fig:coefficients}.
The horizontal dashed line separates the weak from the strong interaction regime.
}
\label{fig:phasediagram}
\end{center}
\end{figure}

As a result of our mean-field approach we have found that the
electron-hole gas in two separate layers has three characteristic
different regimes. According to the phase diagram in Fig.
\ref{fig:phasediagram}, there is a dilute electron-hole gas at weak
interaction without pairing of electrons and holes in the two
layers. In other words, the interaction is too weak to create
pairing, therefore, the electrons and the holes are independent. The
effect of the interaction in this phase is a renormalization of the
Fermi energy, as it can be analyzed by a Hartree-Fock approximation.
Either by increasing the densities due to an increase of the Fermi
energies or by increasing the interaction strength (e.g., by
reducing the interlayer distance $D$) a second order phase
transition to a BCS phase with Cooper pairs appears. Being in the
BCS phase there are two possibilities for changing parameters:
either to increase the densities at a fixed moderate interaction
strength or increase simultaneously the densities and the
interaction strength. In the former case the paired gas undergos
another second order phase transition to a dense electron-hole gas,
whereas in the latter case it is possible that we have a first order
transition from the BCS state to a bosonic Mott state. The Mott
state itself is not directly describable within our mean-field
approach but its existence is indicated by the fact that strongly
coupled electron-hole gas can be represented by a dense hard-core
Bose gas. For the latter the existence of a Mott state has been
established \cite{moseley08}. It is characterized as an
incompressible state of bosons which forms a charge density wave.
The wave vector of the latter is given by the maximal wave vector of
the electron-hole spectrum
$\epsilon_{\mathbf{k},e}+\epsilon_{\mathbf{k},h}$. The phase diagram
in a two-layer electron-hole system can also be characterized with
the help of the Landau expansion in Eq. (\ref{truncated}) by a) weak
interaction at low/high density: $F_2, F_4 >0$,  $\gamma=0$ (EHP),
b) weak interaction at intermediate density: $F_2<0$, $F_4 >0$,
$\gamma>0$ (BCS), c) strong interaction at low/high density:
$F_2>0$, $F_4 <0$, $F_6>0$,  $\gamma=0$ for $E_F>E_b$ (MI), and d)
strong interaction at intermediate density: $F_2>0$, $F_4 <0$,
$F_6>0$, $\gamma>0$ (BCS).


The estimation of the physical parameters, using the results from experimental
measurements on coupled quantum wells in Table \ref{table:1},
indicates that the values $u_0\approx 10^{-4}$ eV, $E_F\approx
10^{-3}$ eV and $k_B T\approx 10^{-4}$ eV are in or close to the
regime of strong interaction and high density. The remaining problem
is the bandwidth $E_b$, which is for typical semiconductors 1 eV.
Therefore, the transition to the Mott state is not accessible in
conventional semiconductors due to the bandwidth being much larger
than the typical Coulomb interaction $u_0$ and Fermi energy $E_F$.
However, it might be possible to create special narrow band
materials and correspondingly large Fermi energies by  appropriate
doping. A possible candidate is a graphene double layer, where the chemical
potential of the electrons and hole can be tuned separately by external
gates. Another advantage of this material is that the mass of the electrons
and the holes can be tuned by a spectral gap in the electronic spectrum \cite{berman12}.
The latter is created by breaking the sublattice symmetry of the honeycomb
lattice. The discovery of new 2D materials might open more opportunities.
Finding the proper material seems to be the main challenge for the experimental
observation of the Mott transition.

\begin{table}
\begin{center}
\begin{tabular}{cccccccc}
material & ${\bar U}$ $[10^{5}\ \mathrm{eV/m}]$ & $q_F^2$ $[10^{-35}\ \mathrm{eV kg}]$ & $R$ & $D$ $[\mathrm{nm}]$  & $k_BT$ $[10^{-4} \ \mathrm{eV}]$ & $a$ $[10^{-25} \ \mathrm{kg m/s}]$ & $r_s$ \\
GaAs/AlGaAs & $1.558$  & $6.542$  & $0.00437$ & $4$  & $1.723$ & $11.14$ & $1.702$  \\
InGaAs & $4.009$  & $17.445$  & $0.00134$  & $4$  & $8.617$ & $4.776$ & $0.551$ \\
ZnCdSe/ZnSe & $2.426$  & $6.542$  & $0.317$ & $4$  & $1.723$ & $154.4$ & $8.751$ \\
\end{tabular}
\caption{Typical experimental values of our mean-field parameters
($m_{0} = 9.10938291 \times 10^{-31} \ \mathrm{kg}$ is the mass of a free electron).
$r_{s} = (\pi n)^{-1/2}/a_{B}$, where $a_{B} =\hbar^{2}\varepsilon/(km_{+}e^{2})$.}
\label{table:1}
\end{center}
\end{table}



\section{Conclusions}

\label{conc}

The BCS-like mean-field approach has provided the phase diagram with
several phases and phase transitions for the electron-hole gas in
two layers with spatially separated electrons and holes. For weak
interaction and low densities we have found a second order
transition from electron-hole plasma to a conventional BCS phase
with electron-hole Cooper pairs. With increasing density there is
another transition from the BCS phase to electron-hole plasma at
weak interaction and a first order transition to a Mott phase at the
strong electron-hole interaction achieved by decreasing the
interlayer separation $D$. At the first order transition from the
BCS phase to the Mott phase, the Fermi energy coincides with the
upper band edge of the electron-hole spectrum. This implies that the
Mott phase is insulating. This behavior can also be understood by
the fact that the Cooper pairs behave like hard core bosons in the
regime of strong interactions. The latter indicates that the Mott
transition of the electron-hole gas is equivalent to the Mott
transition of a hard-core Bose gas at high density. The transition
from the BCS phase to the EHP is a second order phase transition,
whereas the transition to the Mott phase is a first order transition
within our mean field theory. Although it is quite realistic to
observe the BCS phase in the coupled quantum wells and two graphene
layers, the Mott transition is more difficult to access. The main
reason is that it requires a Fermi energy being above the upper band
edge of the electron-hole spectrum. A possible approach could be
based on strong doping of narrow-band semiconductors.


\acknowledgments

The authors acknowledge support from the Center for Theoretical Physics of
the New York City College of Technology, CUNY.


\appendix

\section{First order phase transition}
\label{app:1}

The first order phase transition is characterized by ${\cal
F}_6(\gamma_+)={\cal F}_6(0)$ which also reads
\begin{eqnarray}
F_2\gamma_+^2+F_4\gamma_+^4+F_6\gamma_+^6=0 \ .
\end{eqnarray}
For $\gamma_+>0$ we have two solutions of this quadratic condition as
\begin{eqnarray}
\gamma_+^2=-\frac{F_4}{2F_6}\pm\sqrt{\frac{F_4^2}{4F_6^2}-\frac{F_2}{F_6}}
\ .
\end{eqnarray}
This must be compared with the definition of $\gamma_+^2$ in Eq. (\ref{op2}), which yields
for $\zeta=F_2F_6/F_4^2$ the equations
\begin{eqnarray}
\label{eq}
1\pm 3\sqrt{1-4\zeta}-2\sqrt{1-3\zeta}=0
\ .
\end{eqnarray}
The both Eqs.~(\ref{op2}) and~(\ref{eq}) are solved with $\zeta=1/4$
such that
\begin{eqnarray}
F_4^2=4F_2F_6
\ .
\end{eqnarray}

\end{document}